# TOWARD A MININUM CRITERIA OF MULTI DIMENSIONAL INSTANTON FORMATION FOR CONDENSED MATTER SYSTEMS?

A.W. Beckwith
abeckwith@uh.edu

## ABSTRACT

We present near the end of this document a promising research direction as to how to generalize a technique initially applied to density wave current calculations to questions of instanton formation in multi dimensional condensed matter systems. Initially we review prior calculations done through a numerical simulation that the massive Schwinger model used to formulate solutions to CDW transport in itself is insufficient for transport of soliton-antisoliton (S-S') pairs through a pinning gap model of CDW transport. Using the Peierls condensation energy permits formation of CDW S-S' pairs in wave functionals. This leads us to conclude that if there is a small spacing between soliton-antisoliton (S-S') charge centers, and an approximate fit between a tilted washboard potential and the system we are modeling, that instantons are pertinent to current/transport problems. This requires a very large 'self energy' final value of interaction energy as calculated between positive and negative charged components of soliton-antisoliton (S-S') pairs with Gaussian wave functionals as modeled for multi dimensional systems along the lines of Lu's generalization given below. The links to a saddle point treatment of this instanton formation are make explicit by a comment as to a cosmology variant of instanton formation in multi dimensions we think is , with slight modifications appropriate for condensed matter systems.

# **INTRODUCTION**

1. For the one dimensional case, this is a summary of what can be outlined for this problem.

1a. We used the integral Bogomol'nyi Inequality[1] to present how a soliton-anti soliton (S-S') pair forms

1b Wave functionals are congruent with Lin's nucleation[2] of an electron-positron pair as a sufficiency argument as to forming Gaussian wave functionals[3]

1c We reported compelling reasons as to how additions of a Peierls gap term, in addition to the single chain model resonance condition permits addition of a large potential term which allows for multi chain / Josephson junction treatment as to density wave current calculations[4].

2. How to generalize this for multi dimensional systems?[5]

2a. We speak as to necessary and sufficient conditions for the formation of instantons for certain complex condensed matter system potentials which roughly have to behave similarly as a tilted washboard potential.[6]

2b. In addition the main requirement lies in our upcoming calculations as to formation of a <u>minimum distance</u> requirement between charge centers of a S-S' pair formation used as a nucleus for a Gaussian wave functional representation of an instanton in current calculations.[5]

2c This <u>minimum distance</u> is inversely proportional to the absolute value of a largely electro static 'self energy' value of ions, both positive and negative which interact with each other in a multi dimensional setting.

2d. The false vacuum. hypothesis is a necessary condition for the formation of S-S' pairs and that the multi-chain term we add to a massive Schwinger equation for CDW transport is a sufficiency condition for the explicit formation of a soliton (anti soliton) in our charge density wave transport problem.

Please refer to appendix I, of seven questions as to fundamental issues answered in this document

## **IN ONE DIMENSION WHY WE NEED THE PIERLS GAP TERM ADDED TO OUR PROBLEM?.**

3a. John Miller[6] furthered John Bardeen's work[7] in CDW on a pining gap representation of CDW transport- Use of single chain model for showing how a threshold electric field would initiate transport.~ by use of the extended Schwinger model[6]

$$H = \int_x \left[ \frac{1}{2 \cdot D} \cdot \Pi_x^2 + \frac{1}{2} \cdot (\partial_x \phi_x)^2 + \frac{1}{2} \cdot \mu_E^2 \cdot (\phi_x - \varphi)^2 + \frac{1}{2} \cdot D \cdot \omega_P^2 \cdot (1 - \cos\phi) \right] \quad (3.a)$$

3b Have small driving term $\mu_E \cdot (\phi - \Theta)^2$ added to the main potential term of the washboard potential, is used to model transport phenomenology.

3c. $\mu_E$ is proportional to the electrostatic energy between the S-S' pair constituents (assuming a parallel plate capacitor analogy)

3d $\Theta$ is a small driving force dependent upon a ratio of an applied electric field over a threshold field value. $\Theta = 2 \cdot \pi \cdot \frac{E}{E^*}$ 3e $\Pi_x \equiv D \cdot \partial_t \phi_x$ as canonical momentum density, $D \equiv \left( \dfrac{\mu \cdot h}{4 \cdot \pi \cdot v_F} \right)$,

3f $D \cdot \omega_P^2$ as the pinning energy.

3g $\mu_E$ is electrostatic energy and

3h. Finally, $.01 < \mu_E / D \cdot \omega_P^2 \leq .015$,

## How do we model this sort of problem construction numerically?

4a. Fix quantum mechanically based energy

$$E = i\hbar \frac{\partial}{\partial t} \qquad (4a)$$

4b Fix quantum mechanical momentum

$$\Pi = (\hbar/i) \cdot \partial/\partial \phi \, (x) \qquad (4b)$$

4c Set, $\Theta \equiv \omega_D t$ as a driving force, with $\omega_D$ as a driving frequency. This leads to the following scheme. The first index, j, is with regards to 'space', and the second, n, is with regards to 'time' step. This is Runge-Kutta for the wave functional

$$\phi(j, n+1) = \phi(j, n-1)$$
$$+ i \cdot \Delta t \cdot \left( \dfrac{\hbar}{D} \left[ \dfrac{\phi(j+1,n) - \phi(j-1,n) - 2 \cdot \phi(j,n) + \phi(j+1,n+1) + \phi(j-1,n+1) - 2\phi(j,n+1)}{(\Delta x)^2} \right] - \dfrac{2 \cdot V(j,n)}{\hbar} \phi(j,n) \right) \qquad (4c)$$

4d Also variants of Runge-Kutta in order to obtain a sufficiently large time step interval so as to be $\Delta t \approx 10^{-13}$. Then, the 'massive Schwinger model' is:

$$\phi(j, n+1) = \dfrac{2 \cdot \widetilde{R}}{1 + 2 \cdot \widetilde{R}} \cdot \left( \phi(j-1,n) - \phi(j+1,n) \right) + \dfrac{1 - 2 \cdot \widetilde{R}}{1 + 2 \cdot \widetilde{R}} \cdot \phi(j, n-1)$$

$$- i \cdot \Delta t \dfrac{V(j,n)}{\hbar} \phi(j,n) \qquad (4d)$$

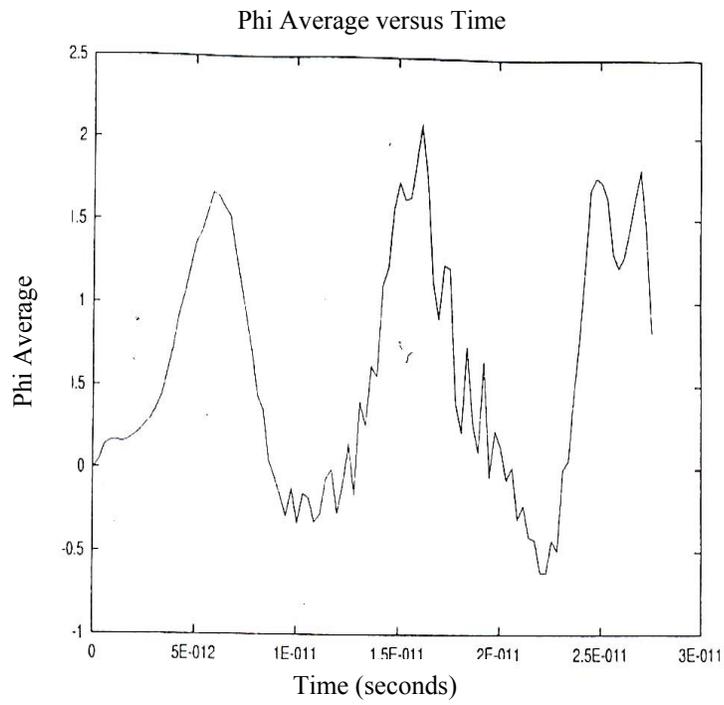

Phi Average versus Time

These numerical schemes lead to **figure 1** above representation of the following type

*Beginning of resonance phenomena due to using the traditional Crank – Nickelson numerical iteration scheme of the one chain model. Phi refers to a time dependent phase value due to a single-chain approximation.*

With $\widetilde{R} = -i \cdot \Delta t \dfrac{h}{2 \cdot D \cdot (\Delta x)^2}$ . One then gets resonance phenomena as represented by Fig. 1.above

**ADDITION OF AN NEW TERM IN THE MASSIVE SCHWINGER EQUATION TO PERMIT FORMATION OF A S-S' PAIR, AND ITS LINKS TO FORMING INSTANTON PHYSICS**.

5a Note that in the argument about the formation of a soliton (anti soliton), that we use a multi-chain simulation Hamiltonian with Peierls condensation energy used to couple adjacent chains (or transverse wave vectors) as represented by[4,5]

$$H = \sum_n \left[ \frac{\Pi_n^2}{2 \cdot D_1} + E_1 [1 - \cos\phi_n] + E_2 (\phi_n - \Theta)^2 + \Delta' \cdot [1 - \cos(\phi_n - \phi_{n-1})] \right]$$ (5.a)

5.b with 'momentum 'we define as[4,5,6]

$$\Pi_n = (h/i) \cdot \frac{\partial}{\partial \phi_n}$$ (5b)

5c We then use a nearest neighbor approximation to use a Lagrangian based calculation of a chain of pendulums coupled by harmonic forces to obtain a differential equation which has a soliton solution .To do this, we write the interaction term in the potential of this problem as

$$\Delta^{'}(1-\cos[\phi_n - \phi_{n-1}]) \to \frac{\Delta^{'}}{2} \cdot [\phi_n - \phi_{n-1}]^2 + \text{very small H.O.T.s.} \tag{5c}$$

5d and then consider a nearest neighbor interaction behavior via

$$V_{n.n.}(\phi) \approx E_1[1-\cos\phi_n] + E_2(\phi_n - \Theta)^2 + \frac{\Delta^{'}}{2} \cdot (\phi_n - \phi_{n-1})^2 \tag{5d}$$

5e Here, we set $\Delta^{'} \gg E_1 \gg E_2$, so then this is leading to a dimensionless Sine–Gordon equation we write as[4,5]

$$\frac{\partial^2 \phi(z,\tau)}{\partial \tau^2} - \frac{\partial^2 \phi(z,\tau)}{\partial z^2} + \sin\phi(z,\tau) = 0 \tag{5e}$$

5f. so that

$$\phi_\pm(z,\tau) = 4 \cdot \arctan\left(\exp\left\{\pm \frac{z + \beta \cdot \tau}{\sqrt{1-\beta^2}}\right\}\right) \tag{5f}$$

I.e.

5g where the value of $\phi_\pm(z,\tau)$ is between $0$ to $2\cdot\pi$ .. phase we call in position space

$$\phi(x) = \pi \cdot [\tanh b(x - x_a) + \tanh b(x_b - x)] \tag{5g}$$

See the following representation of this sort of domain wall

# CDW and its Solitons

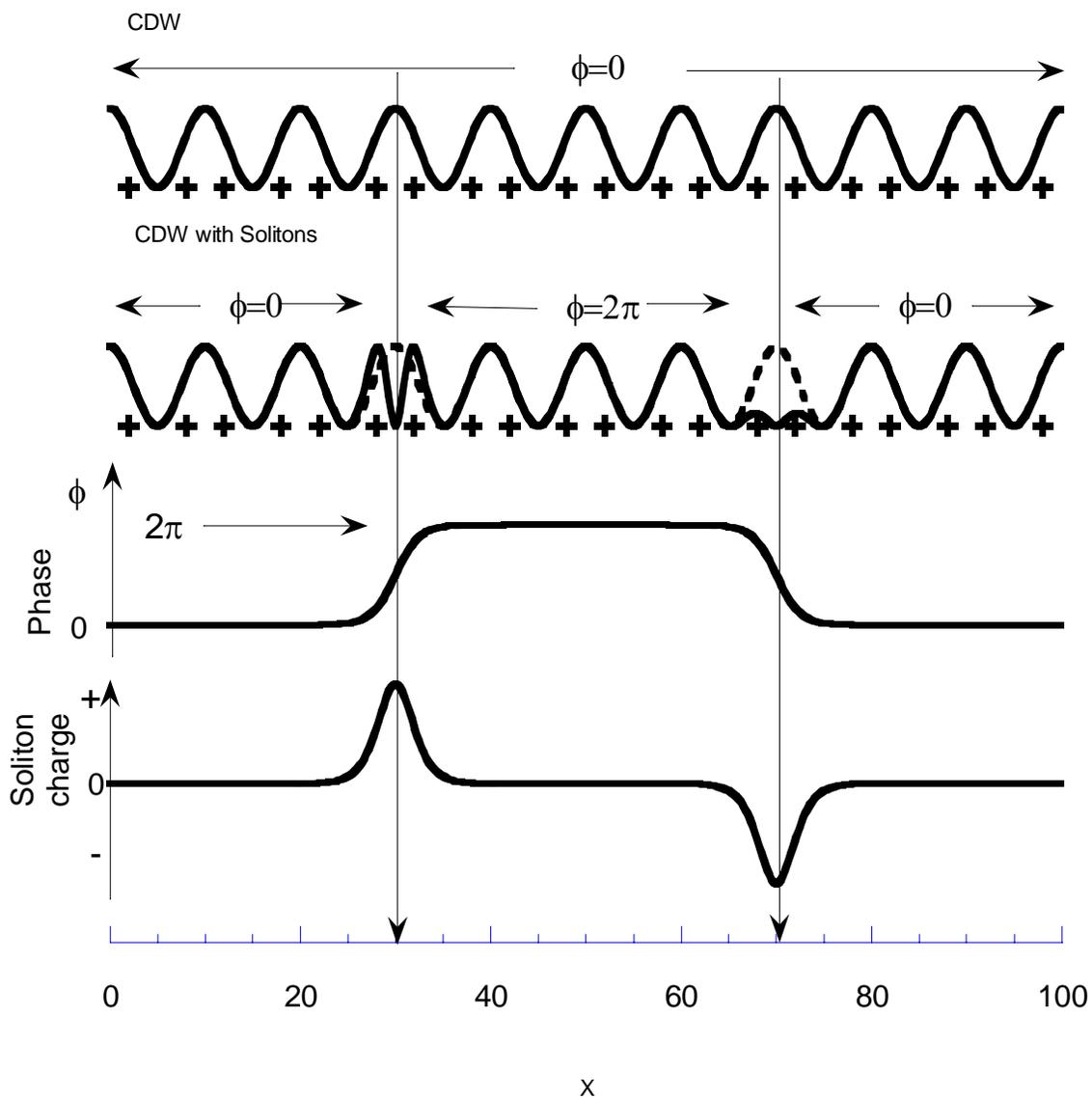

Figure 2

*The above figures represent the formation of soliton-anti soliton pairs along a 'chain'. The evolution of phase is spatially given by*
$\phi(x) = \pi \cdot [\tanh b(x - x_a) + \tanh b(x_b - x)]$

# Now how do we relate all this to the formation of Wave functional Gaussians ?

6a We look at the Lagrangian density $\varsigma$ to having a time independent behavior[4,5]

$$\int d\tau \cdot dx \cdot \varsigma \to t_P \cdot \int dx \cdot L \tag{6a}$$

6b where $t_P$ is the Planck's time interval. Then afterwards, we shall use $\hbar \equiv c \equiv 1$ so we can write

$$\psi \propto c \cdot \exp\left(-\beta \cdot \int L\, dx\right) \tag{6b}$$

6c This was later generalized to be of the form in a momentum space DFT momentum basis in an initial physical state with

$$\alpha \cdot \int dx [\phi_0 - \phi_C]^2_{\phi_C \equiv \phi_T} \equiv \left(\frac{2\cdot\pi}{L}\right)^2 \cdot \sum_n |\phi(k_n)|^2 \tag{6c}$$

6d And a DFT representation of a final state as

$$\alpha \cdot \int dx [\phi_0 - \phi_C]^2_{\phi_C \equiv \phi_F} \equiv \left(\frac{2\cdot\pi}{L}\right)^2 \cdot \sum_n (1 - n_1^2) \cdot |\phi(k_n)|^2 \tag{6d}$$

These in the Charge Density wave case assumed later on that $\phi(k)$ was a momentum space Fourier transform of a soliton-anti soliton pair(S-S') and $n_1 \approx 1 - \varepsilon^+ < 1$ represented the height of this pair reaching its nucleation value, while $\alpha \approx L^{-1}$ was one over the distance between positive and negative charge centers of the S-S' pair.

6e. Also analytical work in momentum space leading to a functional current[4,5]

$$J \propto T_{if} \tag{6e}$$

6f We obtained the final expression which was used[4,5]

$$T_{if} \cong \frac{(\hbar^2 \equiv 1)}{2\cdot m_e} \int \left( \Psi^*_{initial} \frac{\delta^2 \Psi_{final}}{\delta \phi(x)_2} - \Psi_{final} \frac{\delta^2 \Psi^*_{initial}}{\delta \phi(x)_2} \right) \vartheta(\phi(x) - \phi_0(x)) \wp\,\phi(x) \tag{6f}$$

where $\wp\,\phi(x)$ is integration over a variation of paths in the manner of quantum field theory, and $\vartheta(\phi(x) - \phi_0(x))$ is a step function We are assuming quantum fluctuations about the optimum configurations of the field $\phi_F$ and $\phi_T$, while $\phi_0$ represents an intermediate field configuration inside the tunnel barrier

# We pick in both approaches wave functionals

$$c_2 \cdot \exp\left(-\alpha_2 \cdot \int d\tilde{x}[\phi_T]^2\right) \cong \Psi_{final} \tag{7a}$$

And [4,5]

$$c_1 \cdot \exp\left(-\alpha_1 \cdot \int dx[\phi_0 - \phi_F]^2\right) \equiv \Psi_{initial} \tag{7b}$$

with $\phi_0 \equiv \phi_F + \varepsilon^+$ and where $\alpha_2 \cong \alpha_1$

The wave functionals represent the decay of the false vacuum hypothesis, where .

$$\alpha_2 \equiv \Delta E_{gap} \equiv \alpha \approx L^{-1} \tag{7c}$$

We also found that in order to have a Gaussian potential in our wavefunctionals that we needed to have in both interpretations

$$\frac{(\{\ \})}{2} \equiv \Delta E_{gap} \equiv V_E(\phi_F) - V_E(\phi_T) \tag{7d}$$

This is for the following diagram[4,5]

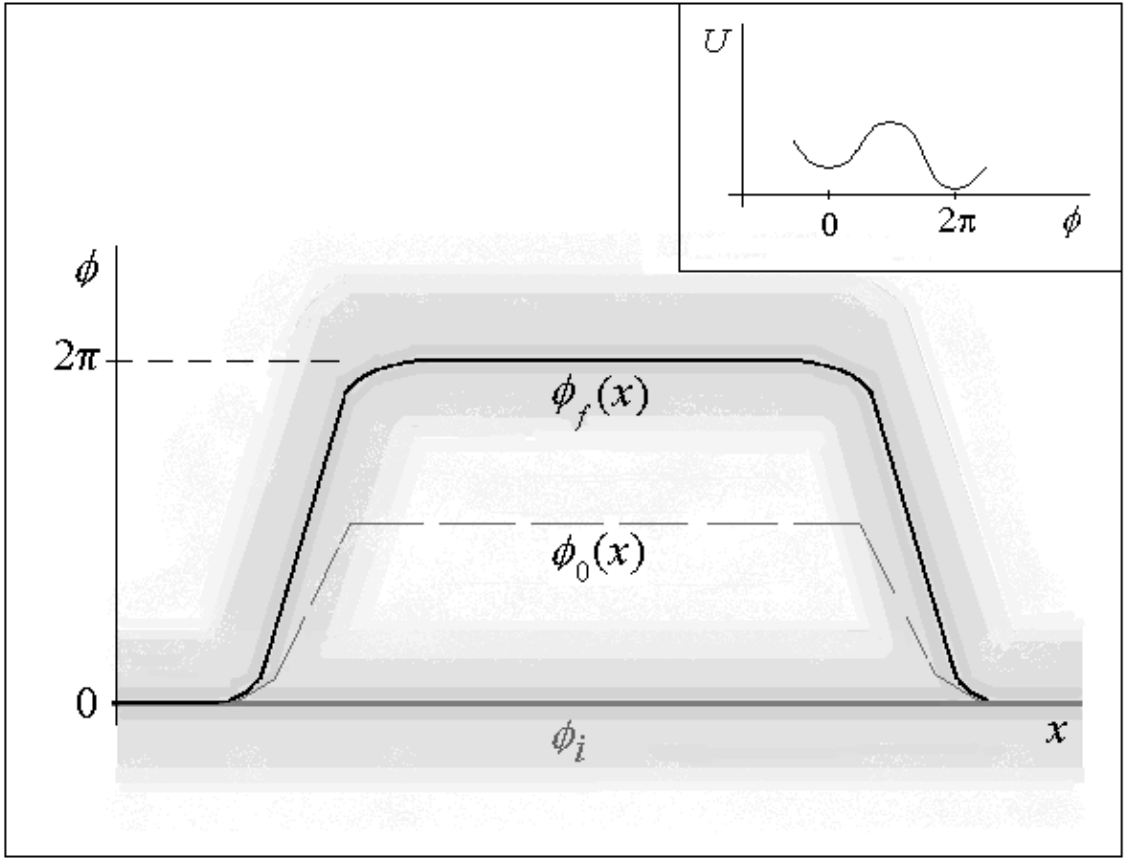

Fig 3

*Evolution from an initial state $\phi_i$ to a final state $\phi_f$ for a double-well potential (inset) in a quasi 1-D model, showing a kink-anti kink pair bounding the nucleated bubble of true vacuum. The shading illustrates quantum fluctuations about the optimum configurations of the field $\phi_F$ and $\phi_T$, while $\phi_0(x)$ represents an intermediate field configuration inside the tunnel barrier. This also shows the direct influence of the Bogomil'nyi inequality in giving a linkage between the 'distance' between constituents of a 'nucleated pair' of S-S' and the $\Delta E$ difference in energy values between $V(\phi_F)$ and $V(\phi_T)$ which allowed us to have a 'Gaussian' representation of evolving nucleated states.*

We had a Lagrangian we modified to be (due to the Bogomil'nyi inequality)[4]

$$L_E \geq |Q| + \frac{1}{2} \cdot (\phi_0 - \phi_C)^2 \cdot \{\ \}$$

(7e)

with topological charge $|Q| \to 0$ and with the Gaussian coefficient found in such a manner as to leave us with wave functionals [1,3,10] we generalized for charge density transport. The end result is seen in [4,5]

$$\Psi_{i,f}\left[\phi(\mathbf{x})\right]\Big|_{\phi \equiv \phi_{ci,cf}} = c_{i,f} \cdot \exp\left\{-\int d\mathbf{x}\, \alpha\left[\phi_{Ci,f}(\mathbf{x}) - \phi_0(\mathbf{x})\right]^2\right\},$$

(7f)

In both cases, we find that the coefficient in front of the wavefunctional in Eq. (3.26) is normalized due to error function integration which lead to the modulus of the tunneling Hamiltonian being proportional to a current which we found was

$$I \propto \tilde{C}_1 \cdot \left[\cosh\left[\sqrt{\frac{2 \cdot E}{E_T \cdot c_V}} - \sqrt{\frac{E_T \cdot c_V}{E}}\right]\right] \cdot \exp\left(-\frac{E_T \cdot c_V}{E}\right)$$

(7g)

This is due to evaluating our tunneling matrix Hamiltonian with the momentum version of an F.T. of the thin wall approximation, [2,3,4,5]

$$\phi(k_n) = \sqrt{\frac{2}{\pi}} \cdot \frac{\sin(k_n L/2)}{k_n}$$

(7h)

The current expression is a great improvement upon the phenomenological Zener current expression, where $G_P$ is the limiting CDW conductance.[4,5]

$$I \propto G_P \cdot (E - E_T) \cdot \exp\left(-\frac{E_T}{E}\right) \quad \text{if } E > E_T$$

(7i)

    0                                otherwise

This is seen in the following figure as given below[2,3,4,5]

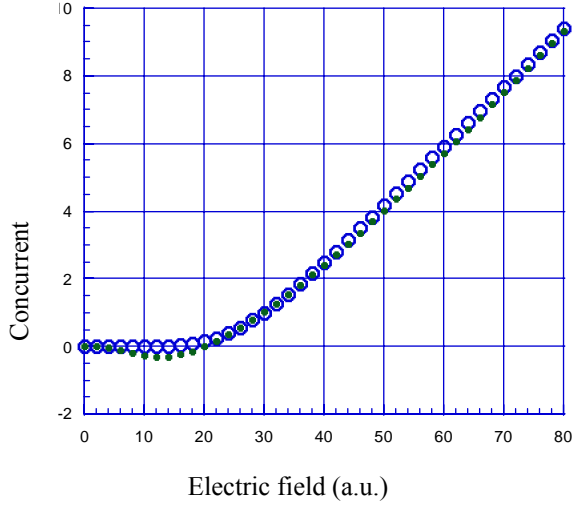

Figure 4

*Experimental and theoretical predictions of current values versus applied electric field. The dots represent a Zenier curve fitting polynomial, whereas the blue circles are for the S-S' transport expression derived with a field theoretic version of a tunneling Hamiltonian. This explains earlier data collected by Miller, Tucker, et al. Also, the classical current gives a negative value for applied electric fields below $E_T$*

## FORMING A CRITERIA FOR NON ZERO MINIMUM DISTANCE BETWEEN S-S' CHARGE PAIRS

Realistically, we need to consider how to include in a calculation as to minimum distance needed for electro static forces needed to insure non zero finite separation between charge centers of a S-S' pair. This is to form a necessary condition for forming an instanton in a condensed matter system.

8a We start off with a representation of the CDW wave, via[5]

$$\Psi \equiv \Psi_0 \cdot \cos(Q \cdot x + \phi(x,t)) \tag{8a}$$

8b As well as setting the phase argument for a thin wall approximation as, initially, at initial time

$$\phi(x, t_{initial}) = 2 \cdot \pi \cdot [\Theta(x - x_s) - \Theta(x_{as} - x)] \tag{8b}$$

This is assuming that $\Theta(x)$ is a typical Heaviside step function..

8c Here, for stationary positive charges, use

$$q_+(x) = \Psi_+ \cos(Q \cdot x) \tag{8c}$$

8d And, for charges which do move, which are negative, we use

$$q_-(x) = \Psi_- \cos(Q \cdot x + \phi(x,t)) \tag{8d}$$

8e So, pick the following expression for total soliton-anti soliton energy, i.e.

$$E_T(L_0) = E_C + E_{S-S'}(L_0) + E_{slide}(L_0) \tag{8e}$$

This has $E_C$ = creation energy for a soliton and an anti soliton separately pair which by experimental results = $\Delta$,

8f i.e. the Pierls gap energy value, and

$$E_C \approx \Delta \tag{8f}$$

8g And if $x_{anti-sol} - x_{sol} = L_0$

$$E_{S-S'}(L_0) = \frac{1}{4 \cdot \pi \cdot \varepsilon_0} \cdot \frac{q^2}{L_0} \tag{8g}$$

As well as, if $U_{slide}^{2 \cdot \pi}$ is the potential energy resulting from changing values of the negative charges due to a shift in the scalar potential $\phi$ over time

$$E_{slide} \equiv L_0 \cdot U_{slide}^{2 \cdot \pi} \tag{8h}$$

8i Then, if we take a partial derivative of the above with respect to $\frac{\partial}{\partial L_0}$ and set the results equal to zero, we find

$$L_0 \equiv \sqrt{\frac{q^2}{4 \cdot \pi \cdot \varepsilon_0} \cdot \frac{1}{U_{slide}^{2 \cdot \pi}}} \tag{8i}$$

8j This is for a situation where we have an electrostatic self energy value of

$$U_{slide}^{2 \cdot \pi} = \frac{1}{4 \cdot \pi \cdot \varepsilon_0} \iint \frac{q_+(x_1) \cdot q_-(x)}{|x_1 - x|} \cdot dx_1 \cdot dx \tag{8j}$$

In the case of CDW, this numerical integral is directly proportional to the energy charge stored between two charged capacitor plates. Leading to a very small value to $L_0$. We need to find ways to generalize this 'self energy' expression above in the case of physical ionic charges in a more generalized multi dimensional setting.

# CONCLUSION: SETTING UP THE FRAMEWORK FOR A FIELD THEORETICAL TREATMENT OF TUNNELING FOR MULTI DIMENSIONAL CONDENSED MATTER SYSTEMS.

Essentially for multi dimensional condensed matter phenomena, we need to observe if or not we a need to generalize what is meant by a wave functional treatment of a multi dimensional Gaussian[8], i.e we find that the Gaussian wavefunctionals would be given in the form given by Lu[9].

9a Lu's integration given below is a two dimensional Gaussian wave functional.

$$|0>^o = N \cdot \exp\left\{ -\int_{x,y} \left[ (\phi_x - \varphi) \cdot f_{xy} \cdot (\phi_y - \varphi) \right] \cdot dx \cdot dy \right\} \tag{9a}$$

Lu's Gaussian wave functional is for a non-perturbed, Hamiltonian as given in Eq. (9b) below

$$H_O = \int_x \left[ \frac{1}{2} \cdot \Pi_x^2 + \frac{1}{2} \cdot (\partial_x \phi_x)^2 + \frac{1}{2} \cdot \mu^2 \cdot (\phi_x - \varphi)^2 - \frac{1}{2} \cdot I_0(\mu) \right] \cdot dx \cdot dy \tag{9b}$$

These two criteria will permit instanton formation in higher dimensional condensed matter system problems. If we look at the one dimensional version of Eqn (9a) above, we have

$$f_{xy} \xrightarrow{reduction-to-one-\dim} \delta(x-y)/L^{1+\delta+} \tag{9c}$$

We have that the distance, L, so noted is a spatial distance between 'instanton' charge pair components , with the very small. coefficient $\delta+$ put in as a power factor denoting a small deviation of this model from purely one dimensional model considerations.

Whereas in multi dimensional treatments, we have

$$f_{xy} \approx \frac{\partial V_{eff}}{\partial r} \tag{9d}$$

If we look at the Bogomolnyi inequality treatment of what happens to the action integral, this gets reduced very quickly to an expression highly dependent upon a Gaussian treatment of this instanton structure in multi dimensions. Our task in the next several months will be to elucidate upon this further.

Furthermore, the seven questions so referenced in the Appendix I entry below are basic issues which will be elaborated upon further in higher dimensions in the next several months

# Appendix I. Seven basic questions for instanton formation, and condensed matter

**1st question**: Why would this be helpful?

*...... involves necessary conditions for formulation of a soliton- anti soliton pair, assuming a minimum distance between charge centers, and discusses the prior density wave physics example as to why a Pierels gap term is added to the tilted washboard potential for insuring the formation of scalar potential fields with an arctan value ranging in value between zero to two pi.*

**Answer**:
The paper so presented is showing necessary and sufficient conditions to find a current which matches experimental data plots in Density wave physics. I.e the following is crucial

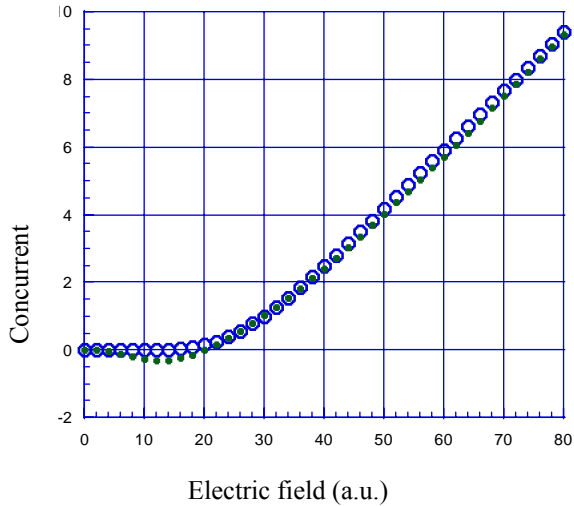

The blue circles represent a current derived of the form

which we found was

$$I \propto \widetilde{C}_1 \cdot \left[ \cosh\left[ \sqrt{\frac{2 \cdot E}{E_T \cdot c_V}} - \sqrt{\frac{E_T \cdot c_V}{E}} \right] \right] \cdot \exp\left( -\frac{E_T \cdot c_V}{E} \right)$$

The current expression is a great improvement upon the phenomenological Zener current expression, where $G_P$ is the limiting CDW conductance, which is the black dots in the above graph

$$I \propto G_P \cdot (E - E_T) \cdot \exp\left( -\frac{E_T}{E} \right) \quad \text{if } E > E_T$$

0  otherwise

This is reflected in the following description as to that figure

*Experimental and theoretical predictions of current values versus applied electric field. The dots represent a Zenier curve fitting polynomial, whereas the blue circles are for the S-S' transport expression derived with a field theoretic version of a tunneling Hamiltonian. This explains earlier data collected by Miller, Tucker, et al. Also, the classical current gives a negative value for applied electric fields below $E_T$*

2$^{nd}$ question.

**Why would one need to worry about Pierls gaps term being added, and tilted band widths in this calculation?**

**Answer** : See the figure below, This was called figure 2 in notes I have

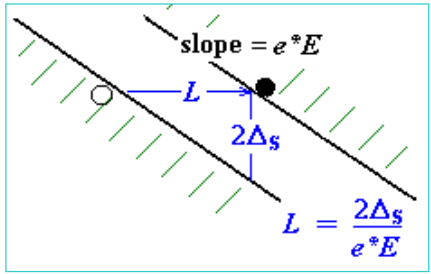

*This is a representation of 'Zener' tunneling through pinning gap with band structure tilted by applied E Field*

The rule for application of the electric field, is as follows

We consider that we will be working with a Hamiltonian of the form

$$H = \int_x \left[ \frac{1}{2 \cdot D} \cdot \Pi_x^2 + \frac{1}{2} \cdot (\partial_x \phi_x)^2 + \frac{1}{2} \cdot \mu_E \cdot (\phi_x - \Theta)^2 + \frac{1}{2} \cdot D \cdot \omega_P^2 \cdot (1 - \cos \phi_x) \right]$$

where the potential system leads to the phenomenology represented in Fig. 2 with what we have been calling $V_E$ the Euclidian action version of the potential given above. In addition, the first term is the conjugate momentum. Specifically, we found that we had $\Pi_x \equiv D \cdot \partial_t \phi_x$ as canonical momentum density, $D \equiv \left( \frac{\mu \cdot h}{4 \cdot \pi \cdot v_F} \right)$, (where $\mu \equiv \frac{M_F}{m_{e^-}} \cong 10^3$ is a Frohlich to electron mass ratio, and $v_F$ is a Fermi velocity $> 10^3$ cm/sec), and $D \cdot \omega_P^2$ as the pinning energy. In addition, we have that $\mu_E$ is electrostatic energy, which is analogous to having a S-S' pair represented by a separation L and of cross-sectional area A, which produces an internal field $E^* = (e^* / \varepsilon \cdot A)$, where $e^* \cong 2 \cdot e^- \equiv$ effective charge and $\varepsilon \equiv 10^8 \cdot \varepsilon_0$ is a huge dielectric constant. Finally, the driving force term, $\Theta = 2 \cdot \pi \cdot \dfrac{E}{E^*}$, where the physics of the term given by $\int dx \cdot \mu_E \cdot (\phi - \Theta)^2$, leads to no instanton tunneling transitions if $\Theta < \pi \Leftrightarrow E < \dfrac{E^*}{2}$ which was the basis of a threshold field of the value $E_T = E^*/2$ due to conservation of energy considerations. Finally, it is important to note that experimental constraints as noted in the device development laboratory lead to $.01 < \mu_E / D \cdot \omega_P^2 \leq .015$, which we claim has also been shown to be necessary due to topological soliton arguments.

### 3$^{rd}$ question :

**What role does the multi chain argument play as far as formulation of the soliton – instanton ? Why add in the Pierls gap term in the first place?**

**Answer**

First of all, we add the following term, based upon the Pierls gap to an analysis of how an instanton evolves

$$H = \sum_n \left[ \dfrac{\Pi_n^2}{2 \cdot D_1} + E_1 [1 - \cos \phi_n] + E_2 (\phi_n - \Theta)^2 + \Delta' \cdot [1 - \cos(\phi_n - \phi_{n-1})] \right]$$

$\Pi_n = (h/i) \cdot \partial / \partial \phi_n$ which then permits us to write

$$U \approx E_1 \cdot \sum_{l=0}^{n+1} [1 - \cos \phi_l] + \dfrac{\Delta'}{2} \cdot \sum_{l=0}^{n} (\phi_{l+1} - \phi_l)^2$$

which allowed using $L = T - U$ a Lagrangian based differential equation of

$$\ddot{\phi}_i - \omega_0^2 [(\phi_{i+1} - \phi_i) - (\phi_i - \phi_{i-1})] + \omega_1^2 \sin \phi_i = 0$$

with

$$\omega_0^2 = \frac{\Delta'}{m_{e^-} l^2}$$

and

$$\omega_1^2 = \frac{E_1}{m_{e^-} l^2}$$

where we assume the chain of pendulums, each of length $l$, leads to a kinetic energy

$$T = \frac{1}{2} \cdot m_{e^-} l^2 \cdot \sum_{j=0}^{n+1} \dot{\phi}_j^2$$

To get this, we make the following approximation.

This has $\Delta'(1 - \cos[\phi_n - \phi_{n-1}]) \to \frac{\Delta'}{2} \cdot [\phi_n - \phi_{n-1}]^2 +$ very small H.O.T.s.

and then consider a nearest neighbor interaction behavior via

$$V_{n.n.}(\phi) \approx E_1[1 - \cos \phi_n] + E_2(\phi_n - \Theta)^2 + \frac{\Delta'}{2} \cdot (\phi_n - \phi_{n-1})^2$$

5e Here, we set $\Delta' \gg E_1 \gg E_2$, so then this is leading to a dimensionless Sine–Gordon equation we write as

$$\frac{\partial^2 \phi(z, \tau)}{\partial \tau^2} - \frac{\partial^2 \phi(z, \tau)}{\partial z^2} + \sin \phi(z, \tau) = 0$$

**Punch line. Without the Pierls term added in, we do not get a Sine Gordon equation. No Instanton formulation.**

**4$^{th}$ question.**

**What happens if we have no Pierls gap term in the potential system ? Can we still have CDW tunneling ?**

**Answer** :

**Nope**. Here is why

Set, $\Theta \equiv \omega_D t$ as a driving force, with $\omega_D$ as a driving frequency. This leads to the following scheme. The first index, j, is with regards to 'space', and the second, n, is with regards to 'time' step. This is Runge-Kutta for the wave functional

$$\phi(j, n+1) = \phi(j, n-1) + i \cdot \Delta t \cdot \left( \frac{\hbar}{D} \left[ \frac{\phi(j+1,n) - \phi(j-1,n) - 2 \cdot \phi(j,n) + \phi(j+1,n+1) + \phi(j-1,n+1) - 2\phi(j,n+1)}{(\Delta x)^2} \right] - \frac{2 \cdot V(j,n)}{\hbar} \phi(j,n) \right)$$

4d Also variants of Runge-Kutta in order to obtain a sufficiently large time step interval so as to be $\Delta t \approx 10^{-13}$. Then, the 'massive Schwinger model' is:

$$\phi(j, n+1) = \frac{2 \cdot \widetilde{R}}{1 + 2 \cdot \widetilde{R}} \cdot (\phi(j-1,n) - \phi(j+1,n)) + \frac{1 - 2 \cdot \widetilde{R}}{1 + 2 \cdot \widetilde{R}} \cdot \phi(j, n-1)$$

$$- i \cdot \Delta t \frac{V(j,n)}{\hbar} \phi(j,n)$$

These numerical schemes lead to **figure 1** below representation of the following type. Bottom line,

No tunneling. Just massive resonance behavior with the wave fuinctional sliding back and forth.

Always

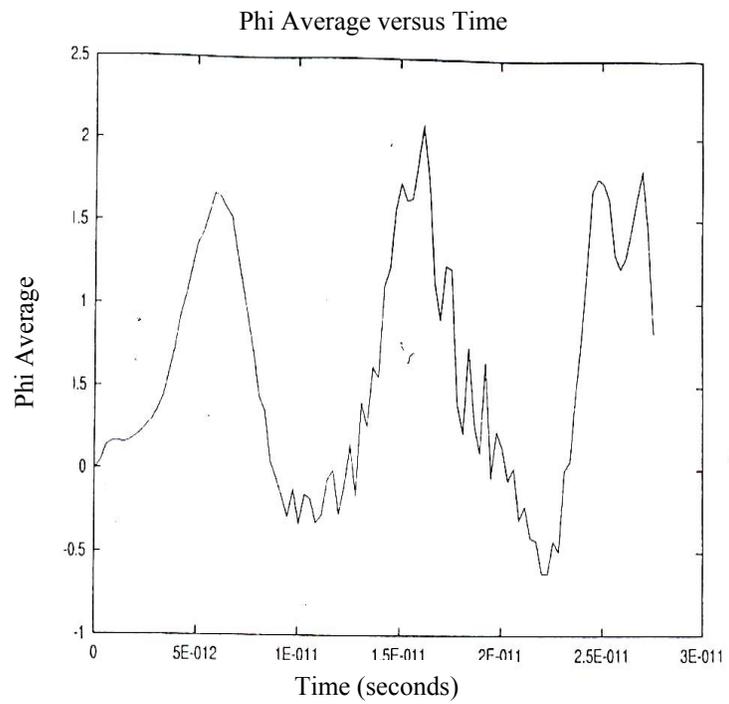

Phi Average versus Time

*Beginning of resonance phenomena due to using the traditional Crank – Nickelson numerical iteration scheme of the one chain model. Phi refers to a time dependent phase value due to a single-chain approximation.*

**5<sup>th</sup> question.**

**How do we represent the wave functional representation of an instanton ?**

**Answer :**

$$\Psi_{i,f}\left[\phi(\mathbf{x})\right]_{\phi \equiv \phi_{ci,cf}} = c_{i,f} \cdot \exp\left\{-\int d\mathbf{x}\, \alpha\left[\phi_{Ci,f}(\mathbf{x}) - \phi_0(\mathbf{x})\right]^2\right\},$$

This ties in with the following false vacuum picture, as elaborated upon by Sidney Coleman

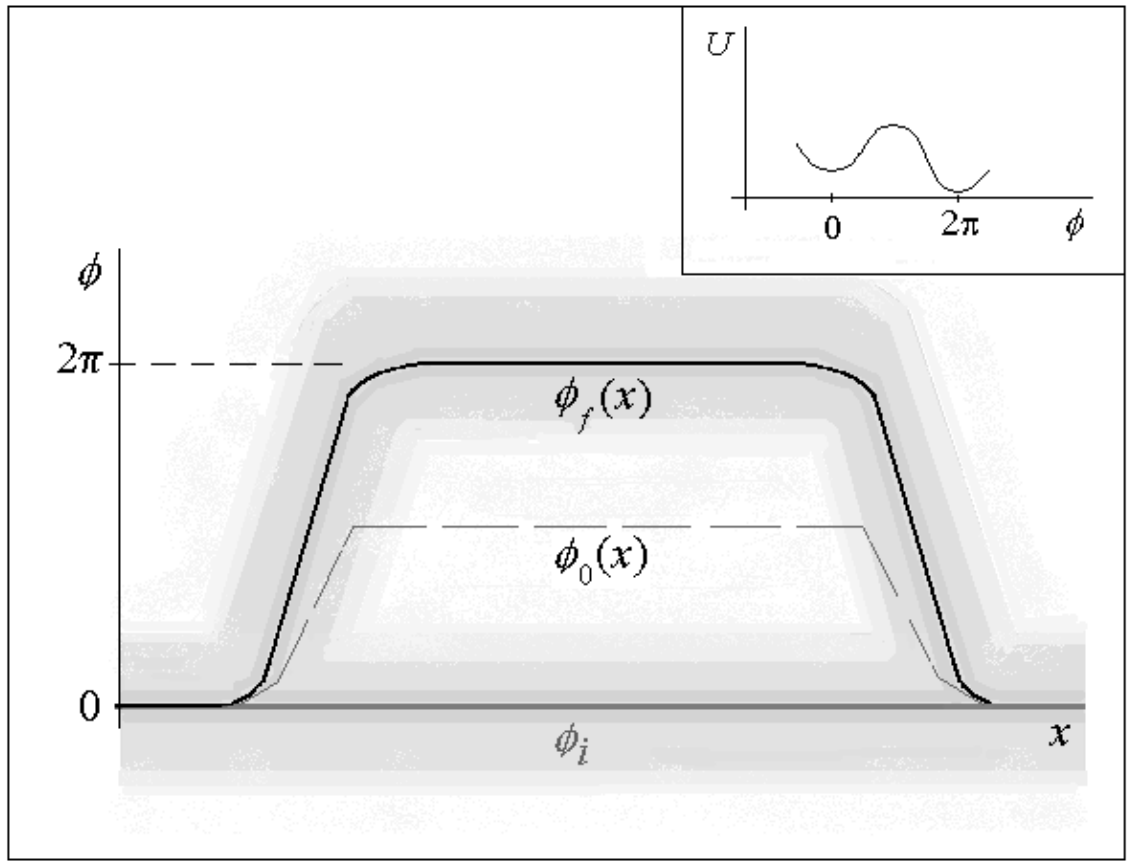

*Evolution from an initial state $\phi_i$ to a final state $\phi_f$ for a double-well potential (inset) in a quasi 1-D model, showing a kink-anti kink pair bounding the nucleated bubble of true vacuum. The shading illustrates quantum fluctuations about the optimum configurations of the field $\phi_F$ and $\phi_T$, while $\phi_0(x)$ represents an intermediate field configuration inside the tunnel barrier. This also shows the direct influence of the Bogomil'nyi inequality in giving a linkage between the 'distance' between constituents of a 'nucleated pair' of S-S' and the $\Delta E$ difference in energy values between $V(\phi_F)$ and $V(\phi_T)$ which allowed us to have a 'Gaussian' representation of evolving nucleated states.*

## 6$^{th}$ question.

**How do we generalize this sort of picture from one dimension to more dimensions ?**

**Answer** :

Lu's integration given below is a two dimensional Gaussian wave functional.

$$|0>^o = N \cdot \exp\left\{-\int_{x,y}\left[(\phi_x - \varphi)\cdot f_{xy} \cdot (\phi_y - \varphi)\right]\cdot dx \cdot dy\right\}$$

Lu's Gaussian wave functional is for a non-perturbed, Hamiltonian as given in Eq. (9b) below

$$H_O = \int_x\left[\frac{1}{2}\cdot \Pi_x^2 + \frac{1}{2}\cdot(\partial_x\phi_x)^2 + \frac{1}{2}\cdot\mu^2\cdot(\phi_x - \varphi)^2 - \frac{1}{2}\cdot I_0(\mu)\right]\cdot dx \cdot dy$$

(9b)

These two criteria will permit instanton formation in higher dimensional condensed matter system problems. If we look at the one dimensional version of Eqn (9a) above, we have

$$f_{xy} \xrightarrow[\text{reduction-to-one-dim}]{} \delta(x-y)/L^{1+\delta+} \tag{9c}$$

We have that the distance, L, so noted is a spatial distance between 'instanton' charge pair components, with the very small. coefficient $\delta+$ put in as a power factor denoting a small deviation of this model from purely one dimensional model considerations.

Whereas in multi dimensional treatments, we have

$$f_{xy} \approx \frac{\partial V_{eff}}{\partial r} \tag{9d}$$

If we look at the Bogomolnyi inequality treatment of what happens to the action integral, this gets reduced very quickly to an expression highly dependent upon a Gaussian treatment of this instanton structure in multi dimensions. Our task in the next several months will be to elucidate upon this further.

### 7th question.

**Minimum criteria for formation of an instanton ? In multi dimensions ?**

**Answer.** It comes from finding a minimum non zero distance between charge centers.

## FORMING A CRITERIA FOR NON ZERO MINIMUM DISTANCE BETWEEN S-S' CHARGE PAIRS

Realistically, we need to consider how to include in a calculation as to minimum distance needed for electro static forces needed to insure non zero finite separation between charge centers of a S-S' pair . This is to form a necessary condition for forming an instanton in a condensed matter system.

8a We start off with a representation of the CDW wave, via[5]

$$\Psi \equiv \Psi_0 \cdot \cos(Q\cdot x + \phi(x,t)) \tag{8a}$$

8b As well as setting the phase argument for a thin wall approximation as, initially, at initial time

$$\phi(x, t_{initial}) = 2 \cdot \pi \cdot [\Theta(x - x_s) - \Theta(x_{as} - x)] \tag{8b}$$

This is assuming that $\Theta(x)$ is a typical Heaviside step function..

8c Here, for stationary positive charges, use

$$q_+(x) = \Psi_+ \cos(Q \cdot x) \tag{8c}$$

8d And, for charges which do move, which are negative, we use

$$q_-(x) = \Psi_- \cos(Q \cdot x + \phi(x,t)) \tag{8d}$$

8e So, pick the following expression for total soliton-anti soliton energy, i.e.

$$E_T(L_0) = E_C + E_{S-S'}(L_0) + E_{slide}(L_0) \tag{8e}$$

This has $E_C$ = creation energy for a soliton and an anti soliton separately pair which by experimental results = $\Delta$,

8f i.e. the Pierls gap energy value, and

$$E_C \approx \Delta \tag{8f}$$

8g And if $x_{anti-sol} - x_{sol} = L_0$

$$E_{S-S'}(L_0) = \frac{1}{4 \cdot \pi \cdot \varepsilon_0} \cdot \frac{q^2}{L_0} \tag{8g}$$

As well as, if $U_{slide}^{2\cdot\pi}$ is the potential energy resulting from changing values of the negative charges due to a shift in the scalar potential $\phi$ over time

$$E_{slide} \equiv L_0 \cdot U_{slide}^{2\cdot\pi} \tag{8h}$$

8i Then, if we take a partial derivative of the above with respect to $\dfrac{\partial}{\partial L_0}$ and set the results equal to zero, we find

$$L_0 \equiv \sqrt{\frac{q^2}{4 \cdot \pi \cdot \varepsilon_0} \cdot \frac{1}{U_{slide}^{2 \cdot \pi}}} \qquad (8i)$$

8j This is for a situation where we have an electrostatic self energy value of

$$U_{slide}^{2 \cdot \pi} = \frac{1}{4 \cdot \pi \cdot \varepsilon_0} \iint \frac{q_+(x_1) \cdot q_-(x)}{|x_1 - x|} \cdot dx_1 \cdot dx$$